\documentstyle[12pt]{article}
\title{HYBRID STATES}
\vspace{2cm}
\author{ Yu.S.Kalashnikova \\ Institute of Theoretical and Experimental
 Physics\\ 117259,
Moscow}
\date{}
\newcommand{\be}{\begin{equation}}
\newcommand{\ee}{\end{equation}}
\begin{document}
\maketitle

\begin{abstract}
 Theoretical arguments are given in favour of existence of gluonic
 degrees of freedom at the constituent level. Models for hybrid
 mesons are discussed, and the predictions are compared with the data
 on meson spectroscopy in the light quark sector. It is demonstrated
 that the $q\bar q g$ content  might be responsible for the
 properties of some recently found mesonic states.
 \end{abstract}

\section{Introduction}

Gluons are present in the QCD Lagrangian on the same footing as
quarks. Nevertheless,"real" theoreticians consider the idea of
constituent glue as a toy for phenomenologists, admitting, on the other
hand, that when the region of large distancesin QCD is under
discussion, the constituent quarks may form rather reasonable and
helpful basis. QCD is highly nonlinear theory, and there is no doubts
that gluons are not only confining, but also are confined, so that
glueballs and hybrids should exist as well as pure $q\bar q$ mesons.
{}From this point of view constituent gluons are nothing but the part of
physically well--motivated constituent basis. At present the direct
QCD studies in the strong coupling regime are not able to define the
appropriate basis in unambiguous way, leaving room for various
 QCD--motivated models.

\section{Wilson loop, area law and surface vibrations}

Confinement is most usually discussed in terms of Wilson loop,
\be
W(C)=S_P ~P~exp~ig \oint_C A^a_{\mu}\lambda_a  dz_{\mu},
\ee
where $\{ \lambda_a\}$ is the octet of colour matrices, and
$P$ stands to order  these matrices along the closed contour $C$.
This celebrated quantity participates in the loop equations [1], enters
in the natural way the Feynman--Schwinger representation
for hadronic Green functions [2], is measurable on the lattice
 [3], and so on.
In this language confinement means the area law for the asymptotically
 large Wilson loop averaged over all gluonic fields:
  \be
  <W(C)>\to N_c exp(-\sigma S).
  \ee

  Here $S=S_{min}$ is the minimal area inside the contour $C$ , and
$\sigma$ is the string tension. The area law is observed in the lattice
simulations of gluodynamics. More precisely, the energy density
distribution is measured between two infinitely heavy colour sources,
with the result corresponding to the area law (2) at large distances
 [3].

If one takes the rectangular contour $R\times T$ with the size $T$ in
the time direction much larger than $R$, the area law gives rise to the
 linear interquark potential
    \be
    V=\sigma R.
    \ee
    On the other hand, the area law corresponds to the effective
    action of the string.

 QCD is surely not the  string theory: at small distances it is the
theory of quarks and gluons interacting perturbatively.
    The existence of area law, nevertheless, means that at large
    distances QCD can be  reduced to  some effective string theory,
    with string tension $\sigma$
    being the
 new universal constant which introduces the new scale and governs
      the behaviour of this effective theory.

 Indeed, let us consider the quenched approximation, with no dynamical
quarks, and, consequently, no possibility for additional quark pair
creation. Then at large interquark distances, $R\gg 1/\sqrt{\sigma}$,
nothing interesting might happen with the string. The only
degrees of freedom in the system are the quark ones, the interaction
energy grows linearly, and we find ourselves in the valent quark sector.

Announcing the string to exist one ought to pay the corresponding price:
 at the interquark
distances $R\sim 1/\sqrt{\sigma}$ the string degrees of freedom start to
reveal themselves. It means that even in the quenched approximation the
spectrum of the system should be much more rich than the spectrum of
the valent $q\bar q$ pair. It is quite natural to identify these extra
excitations with excitations of constituent glue.
In the other words, the gluonic degrees of freedom are responsible for
the string vibrations.

\section{Models, or how to make the QCD string vibrate}

In spite of tantalising efforts the relation between effective
string theory and underlying dynamics of QCD remains rather misty.
Still a lot of heuristic arguments can be given, based on
lattice QCD, strong coupling expansion or stochastic picture
of confinement, in favour of various models, not saying
a word about very naiv\'{e} (but still rather useful) approaches
like bag or potential ones.

The flux tube model is motivated by the strong coupling
expansion. In this model the confining region between the
quark and antiquark is populated with links of flux
that can be extended only in the direction transverse with
 respect to quark--antiquark one. The original
 version of the model assumed the small string oscillations [4];
 later it was shown that small oscillation approximation is inadequate
 [5]. Technically, the string is replaced by the set
of "beads" coupled together by the confining interbead linear force.
 The most elaborated calculations [5] are performed in the "one--bead"
 approximation.

There is no distinguishable gluons in the flux--tube picture, and
phonon--type collective modes play the role of gluonic excitations.
  The conventional $q\bar q$
 meson is the $q \bar q$ pair connected by the string in its
 ground state, while the hybrids are the $q\bar q$ pair
  connected by the excited string.

  Another approach, the constituent string model,
  is motivated by the Vacuum Background Correlators method. It is
assumed that certain background field configurations are responsible
for the confinement [6]. The constituent gluon is introduced as the
 perturbation over confining background [6]. This perturbative
  gluon has nothing to do with the gluon
  of standard perturbation theory. It transforms homogeneously
  under the gauge transformations, and is confined.

  The Green function for a gluon propagating in the given background
  field can be written using the Feynman--Schwinger representation,
  and the Green function for a $q \bar q g$ hybrid is
  constructed averaging the Wilson
  loop operator over the background field confining
   configurations [7]. The $q \bar q$ meson looks
   like a $q \bar  q$ pair connected
by the "minimal" straight--line string, while the $q\bar q g$ hybrid
   is a constituent gluon
   with two straight--line strings, each with quark (or antiquark)
   at the end.

   For the      lowest  states the effective Hamiltonian looks
   like the Hamiltonian of the
   potential model, but the  masses of constituents, including
   the effective gluon mass, are not introduced by hand, but are
   calculated.

   There is a lot of common in the constituent string Hamiltonian
   for a hybrid with one gluon and in the "one--bead"
 flux--tube Hamiltonian, as well as a lot of  numerical differences,
   which appear to compensate each other (for the
   detailed comparison see [8]),
    so that both models end up with similar results for
    the mass of the ground state hybrid with light quarks:
    \be
    M(q\bar q g)=1.7-1.8 GeV
    \ee
    (up to spin--dependent effects).

   The main difference between the models is in quantum numbers. The
    constituent gluon is quite distinguishable, and carries
   the quantum numbers (spin and C--parity) of its own. As the result,
     the possible quantum numbers for the ground state are
     \be
     J^{PC}=0^{\mp +}, ~1^{\mp +}, ~2^{\mp +},~1^{\mp -},
     \ee
 in contrast to the quantum numbers of flux--tube ground state hybrid,
     \be
     J^{PC}=0^{\mp \pm}, ~1^{\mp \pm}, ~2^{\mp \pm},~1^{\mp \mp}.
     \ee

 So, to tell one model from another one should study the $P$--even
hybrids. Unfortunately, all the existing hybrid candidates, as it will
 be demonstrated below, are $P$--odd!

 A lot may be calculated with the described approaches and a lot is
 already calculated, but from the theoretical point of view both
models are still in their infancy. It may appear that these approaches
are not so antagonistic, as it might seem at the first sight. Indeed,
 at present the flux tube knows nothing about vector
     gluons. Nevertheless, attempting to derive the flux tube
     dynamics from QCD one might be forced to introduce
     internal bosonic spin variables populating the string.
As to constituent gluons, they interact with quarks and with each other,
 leading to the $q\bar q -qq\bar g$ mixing. The physical states should
be diagonalized with respect to such mixing, and the resulting string
      configurations might lose their kink--type form and become
     more smooth, resembling the "many--bead" flux tube.

 Having this in mind I will concentrate on $P$--odd hybrids for which
     the expectations from discussed models are similar.

     \section{Hybrids, searched and found (?)}

     For many years as the best signature for hybrids the
     open exotics was considered, namely the exotic quantum numbers.
     The famous $J^{PC}=1^{-+}$ assignment cannot be reached in the
     $q\bar q$ system, while there is no apparent veto for such
quantum numbers in the hadronic channels, e.g. $\pi\eta$ in relative
$P$-- wave. During last  years it was realized that nonexotic quantum
 numbers are also very promising in hybrid searches. The
signature for the hybrid decays were established and studied, and it
was shown that the properties  of excited mesons in the mass range
 1.5-2.0 GeV can be explained with the
     admixture of the constituent glue in the mesonic wave function.

     One of these signatures follows from the
     symmetry of the wave functions of the
     states  involved into the decay. It appears that the ground
     state flux tube hybrid cannot
     decay into two mesons with the same space wave function [9].
     The similar signature exists for a constituent hybrid containing
      the electric gluon ($P$--odd ground states (5) [10]. It means
      that, in spite of phase space considerations, the decay of
      hybrid into two $S$--wave mesons in forbidden, or, at least,
      suppressed, when the decay products do not have {\it a priori}
      the same wave functions, like $\rho\pi$. So the
      hybrids should be looked for
      in the $S$--wave $+P$--wave
       final states, and should be relatively narrow.

 Another, more  sophisticated selection rule is the consequence of
      the decay mechanism. The constituent hybrid decays via
      conversion of a gluon into the $q\bar q$ pair, and the total
      spin of constituents is conserved in  the decay, not only the
      total  angular momentum. The decay of the  flux tube hybrid
      proceeds via string breaking, or creation of additional $q\bar
      q$ pair with $~^3P_0$ quantum numbers, that effectively leads
      to the same spin content of the decay products as for the
constituent hybrid. If the same $^3P_0$ mechanism is responsible for
the decays of conventional $q\bar q$ mesons, then we have rather
powerful tool to distinguish between $q\bar q$'s and $q\bar q g$'s.

In what follows I discuss the resonant activity in the $P$--odd
channels in the mass range 1.5-2.0 GeV, paying attention to the
peculiarities which do not fit $q\bar q$ classification, but may be
explained by hybrid dynamics.

\underline{$1^{-+}$}. The resonant phase motion was reported by BNL
in the $\pi f_1$ final state in the reaction $\pi^-p\to \pi f_1p$ [11]
around the mass of  2.0 GeV. The signal was characterized as broad,
and more statistic is needed. The VES collaboration [12] does not see
this signal in the same reaction (that partly may be explained by the
difference in the incident beam momentum). In  any case, the $\pi
f_1$ is not the main hybrid decay mode, and it is very instructive
to study the $\pi b_1$ final state. In the constituent model the
ratio of partial widths is $$ \pi b_1 : \pi f_1 = 4:1\;.  $$

\noindent
\underline{$1^{--}$}. The vector meson sector is believed to be
well-understood: the $\rho'(1460)$ and $\omega'(1440)$ are considered
as $2~^3 S_1 q \bar{q}$ states, while the upper
$\rho^{''}/\omega^{''}$ are thought of as $~^3D_1 q\bar{q}$'s. The
detailed analysis of multi-pion modes does not support, however, this
assignment [13,14]. There is the additional suppression for the decay
of $2^3 S_1 q\bar{q}$ into $S$-wave + $P$-wave final state because of
the  node in the radial excitation wave function, while the main
decay mode of $\rho'(1460)$ appears to be $\pi a_1$, in accordance
with hybrid dynamics. The admixture of hybrid  is able  to explain
also the decay properties of $\omega'(1440)$, as well as $\rho^{''}$
 and $\omega^{''}$ [14].

\noindent
\underline{$0^{-+}$}. Very clear hybrid signal in the pseudoscalar
sector was reported by VES [12] from the reaction $\pi p \to Hp$, with
the mass about 1800 MeV and width about 200 MeV. This pion decays
strongly into $\pi f_0(1300)$, $\pi f_0(980)$, $KK^*_0$ final states,
with no visible signal in $\pi \rho$ and $KK^*$ channels. Another
argument in favour of hybrid assignment is the value of width: the
state is narrow in contrast to the expectations from the second
radial excitation (the $\pi(1300)$ which is believed to be the first
radially excited pion is very broad, $\Gamma \sim 500$ MeV). The
$\pi(1800)$ decays also into $\pi f_0(1500)$, and $f_0(1500)$ is
treated as the best glueball candidate [15].

\noindent
\underline{$2^{-+}$}. The isovector $2^{-+}$ state around 1.7 GeV,
the $\pi_2(1670)$is considered as $^1 D_2 q\bar{q}$ state, with the
decay properties more or less in agreement with $q\bar{q}$ content.
There is another state $\pi(1770)$, however, seen in the charge
exchange photoproduction [16], for which $J^{PC} = 2^{-+}$ is not
excluded. This state has the width about 100-200 MeV, and decays into
$\pi f_2$, so that this extra state may be a hybrid.

The last but not least is the isoscalar tensor $\eta_2(1870)$ [17]
with the width about 200 MeV and decay modes $\pi a_2$ and $\eta
f_0(980)$.

To conclude, the hybrid candidates start to appear not in single, but
in multiplets, and the emphasis should be moved from pioneering
estimations to dedicated qualitative analysis, both theoretical and
experimental.

I acknowledge the support from International Science Foundation and
Russian Government Grant No. J77100, and Russian Fundamental
Research Foundation, Grant No. 93-02-14937.

\end{document}